\documentclass[reprint,
					aps,
					pra,
					english,
					10pt
]{revtex4-1}

	\usepackage[T1]{fontenc}

	\usepackage[intlimits]{amsmath}

	\usepackage{fixmath}

	\usepackage{amssymb}

	\usepackage{mathtools}
	\mathtoolsset{showonlyrefs}

	\usepackage[caption=false]{subfig}

	\usepackage{dsfont}

	\usepackage{graphicx}
	
	\usepackage[final,colorlinks=true,linkcolor=blue,citecolor=blue]{hyperref}

  \newcommand*{\unity}{\mathds{1}}
  
  \renewcommand*\d[1]{d #1\,}
  \newcommand{\ii}{i}
  \newcommand*{\per}{\operatorname{perm}}
  \newcommand{\abs}[1]{\left\lvert{#1}\right\rvert}
  
  \newcommand*{\defeq}{\mathrel{\vcenter{\baselineskip0.5ex \lineskiplimit0pt
                     \hbox{\scriptsize.}\hbox{\scriptsize.}}}%
                     =}
  \renewcommand{\Omega}{\varOmega}
  \renewcommand*\vec[1]{\mathbold{#1}}
  
  \newcommand{\ket}[1]{\vert #1 \rangle}

  \newcommand{\matrixel}[3]{\langle #1 \vert #2 \vert #3 \rangle}

\begin{document}
  \title{Multiboson Correlation Interferometry with arbitrary single-photon pure states}
  \author{Vincenzo Tamma}
  \email{vincenzo.tamma@uni-ulm.de}
  \author{Simon Laibacher}
  \affiliation{Institut f\"{u}r Quantenphysik and Center for Integrated Quantum
  Science and Technology (IQ\textsuperscript{ST}), Universit\"at Ulm, D-89069 Ulm, Germany}

  \begin{abstract}  
We provide a compact full  description of multiboson correlation measurements of arbitrary order $N$ in
passive linear interferometers with arbitrary input single-photon pure states.
This allows us to physically analyze the novel problem
of \emph{multiboson correlation sampling} at the output of random linear interferometers. 
Our results also describe general multiboson correlation landscapes for an arbitrary number of input single photons and arbitrary interferometers. In particular, we use two different schemes to demonstrate, respectively, arbitrary-order quantum beat interference and $100 \% $ visibility entanglement correlations even for input photons distinguishable in their frequencies. 
  \end{abstract}
  \maketitle

\textbf{\textit{Motivation}}.
Multiboson interference based on correlated measurements is a fundamental phenomenon in atomic, molecular
and optical physics with numerous applications in quantum information processing 
\cite{Pan2012,Knill2001}, quantum metrology \cite{HanburyBrown1956,Motes2015,DAngelo2008}, and imaging \cite{Pittman1995}.
The well-known two-boson interference
``dip'' \cite{Shih1988,Hong1987,Kaufman2014,Lopes2015} is recorded when two single bosons
impinge on a balanced beam splitter and joint detections are performed at the output channels. The dip is a manifestation of the destructive quantum interference between the two-boson quantum
paths corresponding to both bosons being reflected or transmitted. Recent works \cite{Yao2012,Ra2013,Metcalf2013,Broome2013,Crespi2013,tillmann2013experimental,Tillmann2014,Spring2013,Shen2014,Spagnolo2014,Carolan2013,Tichy2014,deGuise2014,Tan2013,Ou2006,Ou2008} have demonstrated the
feasibility of multiboson experiments based on higher-order correlation
measurements well beyond two-boson experiments, which are
crucial towards quantum networks of arbitrary dimensions and the demonstration that boson sampling devices are probably hard to reproduce classically \cite{aaronson2011computational,Franson2013,Ralph2013}.

At the same time, the advent of fast detectors and the production of
single photons with arbitrary temporal and spectral properties
\cite{Keller2004,Kolchin2008,Polycarpou2012} 
make it possible
to fully investigate the temporal dynamics of multiphoton interference via
time-resolving correlation measurements \cite{Tamma2013} by
	using atom-cavity systems \cite{Legero2004}, nitrogen vacancy centers in
diamonds
\cite{Bernien2012,Sipahigil2012}, atomic ions \cite{Maunz2007} and remote
organic
molecules \cite{Wiegner2011}. Two-photon quantum interference as a function of the
detection time  has been observed \cite{Legero2004} in the form
of quantum beats for single photons even when the relative central frequency is
larger than their bandwidths. Moreover, the possibility to encode and retrieve an entire time-dependent quantum alphabet with high fidelity \cite{Nisbet-Jones2013,Monroe2012}
within a given photon spectrum is important for cluster-state quantum computing
\cite{Menicucci2008}, 
 quantum cryptographic schemes \cite{Inoue2002}, and enhanced time metrology \cite{Lamine2008,Pinel2012}.

Finally, higher-order multiphoton interference based on polarization
correlation measurements has been widely used for the generation of
multiqubit entanglement \cite{Greenberger1989,Bouwmeester1999} and tests of
quantum nonlocality \cite{Pan2000}.
This has triggered the implementation of many quantum information applications, including quantum dense coding  protocols \cite{Bennett1992}, 
entanglement swapping, and  teleportation \cite{Bennett1993,Zukowski1993}, entanglement distribution between
distant matter qubits such as ions \cite{Moehring2007} and atomic ensembles \cite{Yuan2008}.

Despite all these remarkable achievements, there is still no
full quantum optical description of time and/or polarization-resolving correlation measurements of arbitrary order in linear multiboson interferometers with input bosons in an arbitrary internal state.
In this letter, we wholly perform such a description and unravel the intimate
connection between the fundamental physics of multiboson interference and 
its computational power.

Although here we consider photonic networks, 
our results are relevant for any interferometric network with bosonic sources, including atoms
\cite{Kaufman2014,Lopes2015}, plasmons \cite{Varro2013} and mesoscopic many-body systems \cite{Urbina2014}, and can be easily extended to Fock states of an arbitrary number of bosons
\cite{TammaLaibacher1} as well as to different input states \cite{TammaLaibacher2,Lund2014,Olson2015}.

\begin{figure} 
  \begin{center}
	  \includegraphics[scale=0.95]{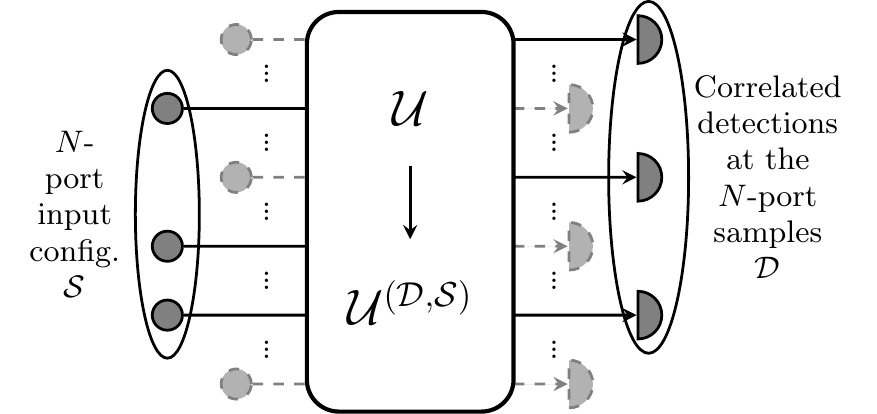}
	  \caption{General setup for multiboson correlation interferometry.  $N$ single bosons are injected into
		  an $N$-port subset $\mathcal{S}$ of the $M$ input ports of a
		  linear interferometer. They can be detected at the output in any possible sample
		  $\mathcal{D}$ containing $N$ of the $M$ output ports at $N$ corresponding detection times
		  $\{t_d\}_{d\in \mathcal{D}}$. For each output port sample $\mathcal{D}$ and given input configuration
		  $\mathcal{S}$, the evolution through the interferometer
		  is fully described by a $N\times N$ submatrix $\mathcal{U}^{(\mathcal{D},\mathcal{S})}$ of the
		  original $M \times
		  M$ interferometer matrix $\mathcal{U}$. The correlated measurements can be performed in any bosonic degree of freedom, such as time, polarization and spin.}
	  \label{fig:InterferometerSetup}
  \end{center}
\end{figure}

  \textbf{\textit{Multiboson Correlation Interferometry}}.  Let us introduce the following general multiphoton correlation experiment based on time- and polarization-resolving measurements
  (see Fig.~\ref{fig:InterferometerSetup}):  $N$ single photons are prepared
  at the $N$ input ports of a linear interferometer with $2M\geq2N$ ports
\footnote{In Ref.~\cite{Reck1994} it was shown that any linear $M$-port interferometer
can be obtained by using a polynomial number (in $M$) of passive linear
optical elements.}.
At the output of the interferometer, we consider all possible correlated
detection events, at given times and polarizations, of the $N$ photons at any $N$-port sample $\mathcal{D}$ of the
$M$ output ports.
The case of boson bunching at the detectors is described in the Supplemental Material.

If $\mathcal{S}$ describes the set of occupied input ports, the $N$-photon input state is
\begin{align}
	\ket{\mathcal{S}} \defeq \bigotimes_{s\in \mathcal{S}}
	\ket{1[\vec{\xi}_{s}]}_{s}
	\bigotimes_{s \notin \mathcal{S}}
	\ket{0}_{s},
	\label{eqn:StateDefinition}
\end{align}
where, using an arbitrary polarization basis $\{\vec{e}_1,\vec{e}_2\}$, the single-photon multimode states are defined as 
\begin{align}
	\ket{1[\vec{\xi}_{s}]}_{s} \defeq \sum_{\lambda=1,2} \int_{0}^{\infty} \d{\omega} \left( \vec{e}_{\lambda}\cdot \vec{\xi}_{s}(\omega) \right) \hat{a}^{\dagger}_{s,\lambda} (\omega) \ket{0}_{s},
	\label{eqn:SinglePhotonState}
	\noeqref{eqn:SinglePhotonState}
\end{align}
with the creation operator $\hat{a}^{\dagger}_{s,\lambda}(\omega)$ for
the frequency mode $\omega$ and the polarization $\lambda$ \cite{Loudon2000}. The direction, magnitude, and phase of the complex spectral distribution $\vec{\xi}_{s}(\omega)$  (with normalization condition $\int
\d{\omega} \abs{\vec{\xi}_{s}(\omega)}^2=1$) define the polarization, the frequency spectrum, and the time of emission of the photon, respectively.

After the evolution in the interferometer, an $N$-photon detection can occur in any $N$-port sample $\mathcal{D}$ at detection times $\{t_d\}_{d\in \mathcal{D}}$ and in the polarizations $\{\vec{p}_{d}\}_{d\in \mathcal{D}}$. For simplicity, we consider input photon spectra in the narrow bandwidth approximation and a polarization-independent interferometric evolution with equal propagation time $\Delta t$ for each possible path.
The field operators $\hat{\vec{E}}^{(+)}_d(t_d)$ at the detected ports $d\in \mathcal{D}$ can then be written in terms of the operators $\hat{\vec{E}}^{(+)}_{s} (t_d-\Delta t)$ at the input ports $s\in \mathcal{S}$ as
\begin{align}
	\hat{\vec{E}}^{(+)}_d(t_d) = \sum_{s\in \mathcal{S}}
	\mathcal{U}_{d,s} \hat{\vec{E}}^{(+)}_s (t_d-\Delta t)
	\label{eqn:ExpansionAout}
\end{align}
through the $N \times N$ submatrix
\begin{align}
	\mathcal{U}^{(\mathcal{D},\mathcal{S})} \defeq [ \mathcal{U}_{d,s} ]_{\substack{d\in \mathcal{D} \\ s\in \mathcal{S}}}
\end{align}
of the $M\times M$ unitary matrix $\mathcal{U}$ describing
the interferometer. 

The rate of an $N$-fold detection event for ideal photodetectors is now given by the $N$th-order Glauber correlation function \cite{Glauber2007}
\begin{equation}
	G^{(\mathcal{D},\mathcal{S})}_{\{t_d,\vec{p}_d\}} \defeq
	\matrixel{\mathcal{S}}{\prod_{d\in \mathcal{D}} \Big(\vec{p}^{*}_d \cdot \hat{\vec{E}}^{(-)}_d(t_d) \Big) \Big( \vec{p}_d \cdot \hat{\vec{E}}^{(+)}_d (t_d) \Big) }{\mathcal{S}}, 
	\label{eqn:ProbabilityRate}
\end{equation}
where $\vec{p}_d\cdot \hat{\vec{E}}^{(+)}_d (t_d)$ is the component of the electric field operator in Eq. \eqref{eqn:ExpansionAout} in the detected polarization $\vec{p}_d$. 

By using the Fourier transforms 
\begin{align}
\vec{\chi}_s(t)\defeq \mathcal{F}[\vec{\xi}_s](t-\Delta t)
	\label{eqn:Fourier}
\end{align}
of the frequency distributions, defining the matrices
\begin{align}
	\mathcal{T}^{(\mathcal{D},\mathcal{S})}_{\{t_d,\vec{p}_d\}} \defeq
	\big[ \mathcal{U}_{d,s} \;\big( \vec{p}_d \cdot \vec{\chi}_s(t_d) \big)
	\big]_{\substack{d\in \mathcal{D} \\ s\in \mathcal{S}}}
		\label{eqn:CorrMatrixDefintion}
\end{align}
and applying the definition of the permanent of a matrix,
\begin{align}
\per \mathcal{M} \defeq \sum_{\sigma \in \Sigma_N} \prod_{i} \mathcal{M}_{i,\sigma(i)},
\label{eqn:PermDef} 
\end{align}
where the sum runs over all permutations $\sigma$ in the symmetric group $\Sigma_N$, the $N$-photon probability rate in Eq.~\eqref{eqn:ProbabilityRate} can be easily expressed 
 as
\begin{align}
	G^{(\mathcal{D},\mathcal{S})}_{\{t_d, \vec{p}_d\}}
	&= \abs{\per \mathcal{T}^{(\mathcal{D},\mathcal{S})}_{\{t_d,\vec{p}_d\}}}^2 
	,
	\label{eqn:CorrelationFinal}
\end{align}
as shown in the Supplemental Material.
Here, the permanent describes the coherent
superposition of $N!$ \textit{detection probability amplitudes} each corresponding to a different \emph{$N$-photon quantum path} from the input ports in $\mathcal{S}$ to the output ports in $\mathcal{D}$. 
Each $N$-photon amplitude is the
product of the $N$ respective single-photon amplitudes, which are
the entries of the matrix $\mathcal{T}^{(\mathcal{D},\mathcal{S})}_{\{t_d,\vec{p}_d\}}$ in Eq.~\eqref{eqn:CorrMatrixDefintion}.
Therefore, the interference between the $N!$ quantum paths depends strongly not only on the interferometric evolution but also on the spectral distributions defining the multiphoton state in Eq.~\eqref{eqn:StateDefinition} and on the detection times and polarizations associated with a measured correlation sample.

\textbf{\textit{Multiboson Correlation Sampling}}. 
The probabilities in Eq.~\eqref{eqn:CorrelationFinal} allow us to physically describe the novel problem of multiboson correlation sampling, i.e. sampling by time and polarization resolving correlation measurements from the probability distribution at the interferometer output.

For approximately equal detection times $t_d\approx t$ and equal polarizations $\vec{p}_d=\vec{p},\, \forall d\in \mathcal{D}$,
the multiphoton detection rate in Eq.~\eqref{eqn:CorrelationFinal} becomes
\begin{align}
	G_{t,\vec{p}}^{(\mathcal{D},\mathcal{S})}
	= \abs{\per \mathcal{U}^{(\mathcal{D},\mathcal{S})}}^2\prod_{s\in \mathcal{S}}  \abs{ \vec{p}\cdot \vec{\chi}_s(t)}^2 .
	\label{eqn:RateIdenticalDets}
\end{align}
which is not trivial if, for each input photon, the detection probability $\abs{\vec{p}\cdot \vec{\chi}_s(t)}^2$ after free propagation is not vanishing at a given time $t$. Interestingly, all $N$-photon quantum paths in Eq.~\eqref{eqn:RateIdenticalDets} are effectively indistinguishable even for non-identical input photons. Their   
interference depends, apart from an overall factor, only on the permanents of submatrices $\mathcal{U}^{(\mathcal{D},\mathcal{S})}$ of the interferometer transformation.
In particular, for random linear interferometers with $30 \lesssim N \ll M$ input photons,
such permanents start to be not
tractable with a classical computer \cite{aaronson2011computational}.
Therefore, the physics of multiboson correlation sampling with non-identical input photons reveals a remarkable potential in quantum information processing \cite{Tamm2015a,Tamm2015,tamma2011factoring}.
\begin{figure}
	\begin{center}
		\subfloat{
			\includegraphics{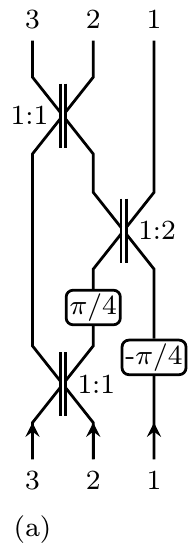}
			\label{fig:QuantumBeatsFirstSetup}
		}
		\subfloat{
			\includegraphics{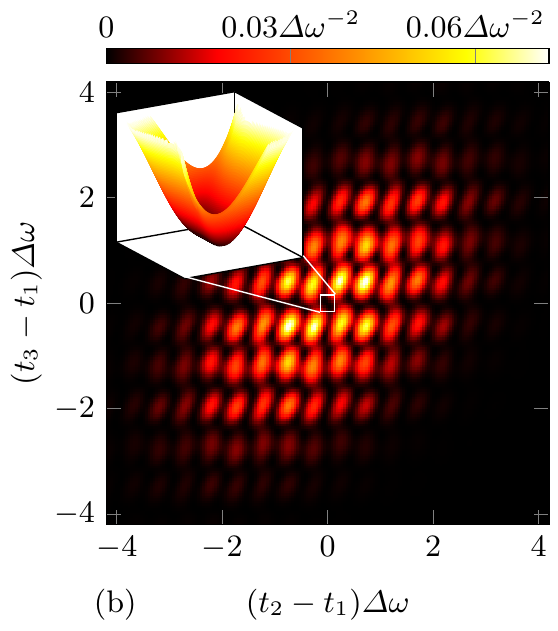}
			\label{fig:QuantumBeatsFirstPlot}
		}

		\subfloat{
			\includegraphics{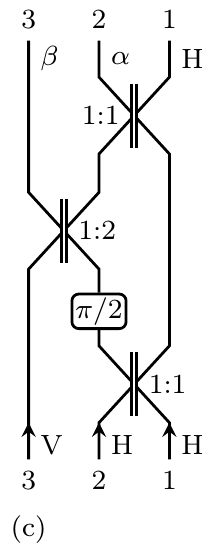}
			\label{fig:QuantumBeatsSecondSetup}
		}
		\subfloat{
			\includegraphics{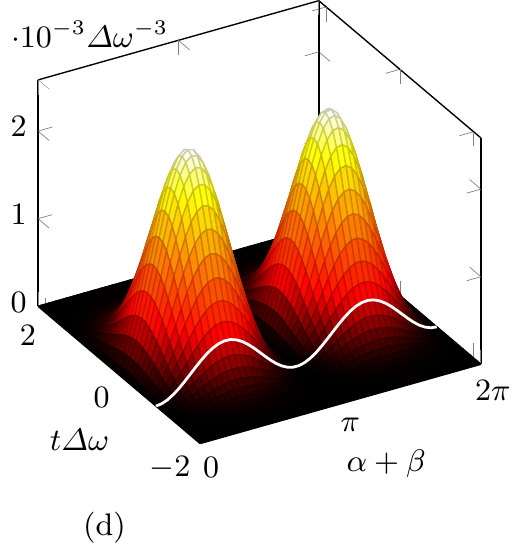}
			\label{fig:QuantumBeatsSecondPlot}
			
		}
		\caption{(Color online) Correlation landscapes at the output of a tritter for three input photons with Gaussian spectra with identical variance
			$\Delta\omega^2$ but different central frequencies
			($\omega_{0,2}-\omega_{0,1}=8.0\Delta\omega$,
			$\omega_{0,3}-\omega_{0,1}=12.7\Delta\omega$). In panel (b),
		we observe three-photon quantum beats in the threefold coincidence rate measured at the output of
	the tritter in panel (a) for equally polarized input photons and polarization-independent detections. The inset magnifies the three-dimensional ``dip'' occurring at equal detection times.
In panel (c), $H$-,$H$-, and $V$-polarized input photons impinge on a symmetric tritter. The emergence of W-state-type correlations for joint detections at approximately equal times $t$ is shown in panel (d) (here, for simplicity, the propagation time is $\Delta t \cong 0$): if a $H$-polarized photon is detected in one output port we observe in the remaining two ports $100 \%$ visibility correlations in the sum $\alpha+\beta$ of the detected polarization angles typical of a Bell state (indicated for an arbitrary time $t$ by the white curve).} 
		\label{fig:Landscapes}
	\end{center}
\end{figure}

\textbf{\textit{Multiboson correlation landscapes}}. 
The general result obtained in Eq.~\eqref{eqn:CorrelationFinal} allows us also to
describe the possible multiboson interference ``landscapes'' 
which arise from correlation measurements in given degrees of freedom (time, polarization, spin, etc.) depending on the internal state of the input bosons
and on the interferometer transformation. As an example, we consider \mbox{$N=3$-photon} correlation measurements in the extreme case of three completely
distinguishable Gaussian single-photon pulses (with identical variances
$\Delta\omega^2$ and relative central frequencies $\omega_{0,2} - \omega_{0,1}
=8.0 \Delta\omega$, $\omega_{0,3} - \omega_{0,1}= 12.7
\Delta\omega$) impinging in two different
$2M=6$-port interferometers.

\textit{Quantum beats}. The first interferometer, shown in Fig.~\ref{fig:QuantumBeatsFirstSetup}, is a tritter characterized by a unitary transformation
\begin{align}
\mathcal{U}=\frac{1}{\sqrt{3}} \left[ \begin{array}{ccc} 1 & \ii & -\ii \\ \ii & (1-\sqrt{3})/2 & -(1+\sqrt{3})/2 \\ \ii & (\sqrt{3}+1)/2 & (\sqrt{3}-1)/2 \end{array} \right] \nonumber
\end{align}
with permanent zero, implying, for three input photons with equal spectra and polarization, a vanishing three-fold
coincidence rate independent of the detection times. Does completely destructive interference occur also for input photons distinguishable in their frequencies? Remarkably, the answer is yes at approximately equal detection times according to Eq.~\eqref{eqn:RateIdenticalDets}. Indeed, the multiphoton landscape depicted  in Fig.~\ref{fig:QuantumBeatsFirstPlot}, corresponding to the three-photon detection rate in Eq.~\eqref{eqn:CorrelationFinal} as a function of the relative detection times $t_2-t_1$ and $t_3-t_2$ for polarization-independent detections, reveals a three-dimensional ``dip''  at the origin. Although the photons are fully distinguishable in their frequencies, the three-photon quantum paths for equal detection times are completely indistinguishable and thereby interfere. More interestingly, departing from the dip at $t_1 \cong t_2 \cong t_3$ we observe three-dimensional quantum beats in the two relative detection times.
These beats emerge from the superposition according to the interferometric evolution of several (in general of the order of $(N!)^2$) interfering terms in Eq.~\eqref{eqn:CorrelationFinal} oscillating with periodicity determined by the frequency differences of the input photons. The beats show a Gaussian damping along both diagonals with a width characterized by the photon coherence time $\Delta\omega^{-1}$.
In general, a plethora of $N$-photon interference landscapes can be obtained by tuning different physical parameters, such as the input internal states, the interferometer evolution and the measurement observables. The emerging multiphoton interference landscapes are thereby a powerful tool to extract information about all these parameters simultaneously.

\textit{Entanglement correlations}. The second interferometer is a symmetric tritter described by the unitary transformation 
\begin{align}
\mathcal{U}=\frac{1}{\sqrt{3}}[\exp(\ii \frac{2\pi}{3}d\cdot s)]_{\substack{d=1,2,3\\s=1,2,3}}
\end{align}
(see Fig.~\ref{fig:QuantumBeatsSecondSetup}) with two input photons horizontally polarized and the third vertically polarized. For input photons identical in their frequency-temporal spectra, a three-fold coincidence measurement would be only sensitive to the entangled state 
\begin{align}
	\ket{W} = \frac{1}{\sqrt{3}}(\ket{H,H,V} + \ket{H,V,H} + \ket{V,H,H})
\end{align}
(so called $W$ state \cite{Duer2000}) independently of the detection time. This is evidently not the case for input photons with different spectral distributions, which are relevant from an experimental point of view. However, here we demonstrate $100 \%$ visibility $W$-state correlations even for input photons completely distinguishable in their frequencies. The emergence of such entanglement correlations is shown in Fig.~\ref{fig:QuantumBeatsSecondPlot} for joint detections at approximately equal times $t$ according to Eq.~\eqref{eqn:RateIdenticalDets}: if an $H$-polarized photon is detected in one output port, we observe in the remaining two ports, at any time $t$, correlations typical of a Bell state in the sum of the detected polarization angles. These correlations arise from the physics of time-resolving correlation measurements: at approximately equal detection times all the multiphoton detection amplitudes fully interfere even for input photons distinguishable in their frequencies. Moreover, similarly to Ref.~\cite{Ramelow2009}, this scheme has the potential to implement more general entanglement correlations both in frquency/time and polarization, with the advantage of not requiring entangled sources.

\textbf{\textit{Non-resolving correlation measurements}}. We now consider the case of correlation measurements which do not resolve the detection times and polarizations, resulting in an average over these degrees of freedom. In this case, we obtain the probability
\begin{align}
	P_{\text{av}}(\mathcal{D};\mathcal{S}) \defeq
	\sum_{\{\vec{p}_d\} \in \{ \vec{e}_1, \vec{e}_2 \}^{\otimes N}} \int_{-\infty}^{\infty} \Big(
	\prod_{d\in \mathcal{D}} \d{t_d} \Big) G^{(\mathcal{D},\mathcal{S})}_{\{t_d,\vec{p}_d\}}
	\label{eqn:IntegratedNotExplicit}
	\noeqref{eqn:IntegratedNotExplicit}
\end{align}
to detect the $N$ photons coming from the input ports $\mathcal{S}$ in the output
ports $\mathcal{D}$, where $\left\{ \vec{e}_1,\vec{e}_2 \right\}$ is an arbitrary polarization basis.

As we show in the Supplemental Material, by defining the \textit{overlap factors}
\begin{align}
	f_{\rho}(\mathcal{S}) &\defeq
	\prod_{s\in \mathcal{S}} \int_{-\infty}^{\infty} \d{t}
	\vec{\chi}_s(t)\cdot \vec{\chi}_{\rho(s)}(t) 
	\label{eqn:DistinguishabilityFactor}
\end{align}
for the interfering $N$-photon detection amplitudes in Eq.~\eqref{eqn:CorrelationFinal} and the \textit{interference-type matrices}
\begin{align}
	\mathcal{A}_{\rho}^{(\mathcal{D},\mathcal{S})}\defeq
	[ \mathcal{U}_{d,s}^{*} \mathcal{U}_{d,\rho(s)} ]_{\substack{d\in \mathcal{D} \\
	s\in \mathcal{S}}},
	\label{eqn:Amatrix}
	\noeqref{eqn:Amatrix}
\end{align}
the probability of an $N$-fold detection in the sample $\mathcal{D}$ can be expressed concisely as
\begin{align}
	P_{\text{av}}(\mathcal{D};\mathcal{S})
	= \sum_{\rho \in \Sigma_N} f_{\rho}(\mathcal{S}) \per \mathcal{A}_{\rho}^{(\mathcal{D},\mathcal{S})}.
		\label{eqn:IntegratedGeneral} 
\end{align}

The probability for each configuration $(\mathcal{D};\mathcal{S})$ in
Eq.~\eqref{eqn:IntegratedGeneral} describes the multiphoton interference in a boson
sampling device and represents a generalization of the
two-photon ``dip'' interference \cite{Hong1987,Shih1988} to a general number
$N$ of single photons in
a linear interferometer with $2M\geq 2N$ ports
\footnote{While completing this manuscript, we came across different
approaches (see Refs.
\cite{Shchesnovich2015},\cite{Rohde2015} and \cite{Tichy2015} for details) from the
one leading to Eq.~(\ref{eqn:IntegratedGeneral}) to describe the so-called boson sampling problem (BSP) \citep{aaronson2011computational}  for arbitrary spectral distributions of the input bosons. Differently
from these approaches, our analysis highlights the dependence
of the $N$-boson sampling probability rates (see Eq.~(\ref{eqn:CorrelationFinal})) on the experimental
detection times and polarizations, which is at the heart of the physics of the more fundamental task of multiboson correlation sampling, as already described.
Furthermore, the experimental emergence of the $N$-boson sampling probabilities (see Eq.~(\ref{eqn:IntegratedGeneral})) 
averaging over all possible detection times allows us to fully physically describe how the interference between all the possible $N!$ multiboson quantum paths depends on the $N$-boson  distinguishability at the detectors.
}. 
If we assume Gaussian temporal distributions $\vec{\chi}_s(t)$ which only differ by a time shift, this result reduces to the one obtained for $N=3$ in
Refs.~\cite{Tan2013,deGuise2014} which relies on the use of immanants
\cite{littlewood1934group}.
Differently from \cite{Tan2013,deGuise2014},
our result is valid for any value $N$, for any single-photon spectra, and depends only on 
``multiphoton interference'' permanents.

We now consider two limiting scenarios:
\paragraph{Absence of $N$-boson interference:}

All $N$-photon quantum paths are distinguishable, corresponding to overlap factors in Eq. \eqref{eqn:DistinguishabilityFactor} $f_{\rho}(\mathcal{S}) \approx 0 \ \forall \rho\neq\unity$.  Therefore,
		the probability in
		Eq.~\eqref{eqn:IntegratedGeneral} is given by the completely
		incoherent superposition
\begin{align}
	P_{\text{av}}(\mathcal{D};\mathcal{S}) &\approx \per \mathcal{A}^{(\mathcal{D},\mathcal{S})}_{\rho=\unity}
\end{align}
with the non-negative matrix
$\mathcal{A}^{(\mathcal{D},\mathcal{S})}_{\rho=\unity} = [ \abs{\mathcal{U}_{d,s}}^2 ]_{\substack{d\in \mathcal{D} \\ s\in \mathcal{S}}}$,
whose permanent can be efficiently estimated \cite{Jerrum2004}. Since, in this case, no multiphoton interference occurs the problem is computationally feasible.
\paragraph{Complete $N$-boson interference:}
All  $N!$  $N$-photon quantum paths are indistinguishable, $f_{\rho}(\mathcal{S})=1 \ \forall \rho$, and interfere. Thereby,
Eq.~\eqref{eqn:IntegratedGeneral} reduces to
\begin{align*}
	P_{\text{av}}(\mathcal{D};\mathcal{S}) & \approx 
	\sum_{\rho \in \Sigma_N} \per \mathcal{A}_{\rho}^{(\mathcal{D},\mathcal{S})} 
	= \abs{\per \mathcal{U}^{(\mathcal{D},\mathcal{S})}}^2.
\end{align*}
The fact that only in this case the output probabilities are determined by  permanents of complex matrices is at the heart of the demonstration of the complexity of boson sampling devices based on non-resolving correlation measurements given in Ref. \cite{aaronson2011computational}.

In the two limits considered, we recover the well-known results \cite{Scheel2005,Metcalf2013,Broome2013} describing the detection probabilities for full multiboson distinguishability and indistinguishability.
In addition, the general result in Eq.~\eqref{eqn:IntegratedGeneral} allows us to fully describe all possible experimental scenarios of partial multiphoton distinguishability. This description triggers exciting questions about the complexity of these scenarios from an experimental point of view.

\textbf{\textit{Discussion}}. 
	We provided a compact full description of multiphoton interferometry based on correlated measurements in time and polarization of any order for arbitrary states of the input photons.

We have physically analyzed the novel problem of multiboson correlation sampling at the output of random linear interferometers.
This is fundamental towards a deeper understanding of the full potential of multiboson quantum interference in quantum information processing in the case of non-identical photons, which is of interest from an experimental point of view.

Moreover, we demonstrated how multiphoton correlation measurements lead to arbitrary-order multiphoton landscapes, which can be tuned with respect to different physical parameters, such as the input internal states, the interferometer evolution and the measured physical observables. These results pave the way for the use of multiphoton interference as a powerful tool for the characterization of  the spectral distribution \cite{Legero2006} of an arbitrary number of single photons and their distinguishability \cite{Lang2013} after the interferometric evolution, which is essential in multiphoton quantum networks \cite{Pan2000}. 

We also showed that even with non-identical input photons it is possible to achieve entanglement correlations with $100 \% $ visibility for an arbitrary number of photons.  This result may lead to real-world applications in quantum information processing with non-identical photons, such as sampling of bosonic qubits \cite{Tamma2014},
nondeterministic nonlinear gates \cite{Knill2001,Rohde2011,Metz2008},
entanglement of an arbitrary number of distant qubits \cite{Legero2004,Sipahigil2012,Halder2007},  time-bin qubit networks \cite{Nisbet-Jones2013,Monroe2012}, quantum teleportation \cite{Gao2013} and quantum communication protocols
\cite{Halder2008}. 

Finally, we provided a full description of arbitrary-order interferometry based on correlation measurements not sensitive to the detected polarizations and times. This description can be applied to the optimization of multiphoton metrology schemes \cite{Motes2015} with non-identical single-photon sources for applications in biomedical physics \cite{Crespi2012a}.

\begin{acknowledgments}
V.T. would like to thank  M. Freyberger, F. N\"{a}gele, W. P.
Schleich, and K. Vogel, as well as J. Franson, S. Lomonaco, T. Pittmann,  and
Y.H. Shih for fruitful discussions during his visit at UMBC in the
summer of 2013.

V.T. acknowledges the support of the German Space Agency DLR with funds
provided by the Federal Ministry of Economics and Technology (BMWi) under
grant no. DLR 50 WM 1136.

This work was supported by a grant from the Ministry of Science, Research and the Arts of Baden-W\"urttemberg (Az: 33-7533-30-10/19/2).
\end{acknowledgments}

\end{document}